# Optical Coherence Tomography in Soft Matter †

Kasra Amini[a,‡], Cornelius Wittig[a,‡], Sofia Saoncella[a], Outi Tammisola[b], Fredrik Lundell[a], and Shervin Bagheri[a]

Optical Coherence Tomography (OCT) has become an indispensable tool for investigating mesoscopic features in soft matter and fluid mechanics. Its ability to provide high-resolution, non-invasive measurements in both spatial and temporal domains bridges critical gaps in experimental instrumentation, enabling the study of complex, confined, and dynamic systems. This review serves as both an introduction to OCT and a practical guide for researchers seeking to adopt this technology. A set of tutorials, complemented by Python scripts, are provided for both intensity- and Doppler-based techniques. The versatility of OCT is illustrated through case studies, including time-resolved velocimetry, particle-based velocity measurements, slip velocity characterization, detection of shear-induced structures, and analysis of fluid-fluid and fluid-structure interactions. Drawing on our experiences, we also present a set of practical guidelines for avoiding common pitfalls.

## 1 Introduction

Many experiments in soft matter involve physics that contain multiple scales in both time and space, which make it particularly challenging to obtain well-resolved data. Optical coherence tomography (OCT) is an attractive option for diagnostics as it allows obtaining high spatial (micrometer) and temporal (100 kHz) resolution three-dimensional images within scattering media. It also allows the acquisition of velocity measurements of fluid flows.

OCT is well-established in medical imaging[1–5] and has increasingly been used in rheology[6–9], profilometry and fluid mechanics[10]. The device is now rapidly expanding into soft-matter applications for several reasons. Firstly, OCT provides tomographic images and, therefore, quantitative spatial information of complex materials. In its most basic form, it works by splitting light into a reference arm and a sample arm; the reflected light from both is recombined to create interference patterns. These patterns, analyzed as a function of depth, provide detailed information about the material's microstructure[11]. Its resolution ($1-10\mu m$) and penetration depth ($\sim 1\,mm$) are larger than confocal microscopy ($< 1\mu m$, $100\mu m$) but smaller than techniques such as ultrasound ($\sim 150\mu m$, $\sim 1\,cm$), high-resolution computed tomography ($\sim 300\mu m$, $\sim 10\,cm$) and magnetic resonance imaging ($\sim 1\,mm$, $> 10\,cm$).

Secondly, OCT has a Doppler feature (D-OCT), which allows for non-intrusive velocimetry in opaque media. The opaqueness inherent in many soft matter systems containing colloids, gels, polymers, cells and organisms poses significant challenges for methods like Particle Image Velocimetry (PIV) and Particle Tracking Velocimetry (PTV), particularly in dynamic conditions and confined geometries. D-OCT works by analyzing the interference between back-scattered light from moving particles and a reference beam. When this interaction occurs, a frequency shift is produced, known as the Doppler frequency, which is proportional to the velocity of the moving sample. D-OCT captures the velocity component that is parallel to the incident light beam[11]. The device has been used to analyze flow patterns in confined geometries, such as T-junctions[12] and flow-focusing devices[12]. It has also resulted in the detailed investigations of phenomena like flow-induced structuring, mixing, and droplet formation in multiphase systems[13–15]. The method also offers a sensitive means of visualizing and quantifying blood flow, especially in small vessels like capillaries[16].

Various forms of OCT devices have had a significant impact in characterizing complex (i.e., non-Newtonian) fluids. For example, Haavisto et al.[6,7] characterized microfibrillated cellulose (MFC) flow and wall slippage in both pipes and rheometric devices, shedding light on flow instabilities and boundary effects that could be captured with OCT. Jalaal et al.[9] employed OCT to investigate the spatio-temporal pattern of gel formation within droplets, providing a unique perspective on in-situ phase changes in fluids. By tracing particles through OCT intensity signals, the authors demonstrated the adequate speed and precision of OCT in visualizing these changes, setting it apart from traditional confocal imaging systems, which required multiple planar images for cross-sectional analysis. Gowda et al.[12] investigated three-dimensional thread-formation under the influence of effective interfacial tension in a colloidal dispersion system in a flow focusing channel. Using D-OCT in combination with numerical simulations, they could capture various flow patterns such as threading, jetting, and dripping. More recently, Jäsberg et al.[17] introduced polarization-sensitive OCT (PS-OCT) as a lightweight tool for observing particle orientation during the processing of cellulose nanocrystals (CNCs). This study demonstrated how PS-OCT could provide real-time online measurements, allowing researchers to track the orientation of non-spherical, elongated particles.

[a] FLOW, Dept. of Engineering Mechanics, KTH, Stockholm SE-100 44, Sweden
[b] FLOW and SeRC (Swedish e-Science Research Centre), Dept. of Engineering Mechanics, KTH, Stockholm SE-100 44, Sweden
† Electronic Supplementary Information (ESI) available: [details of any supplementary information available should be included here]. See DOI: 10.1039/cXsm00000x/
‡ KA and CW had equal contributions.
(Email address for correspondence: kasraa@kth.se / wittig@kth.se / shervinb@kth.se)

OCT has played a crucial role in the field of biofilms[18–20], in particular for the in-situ imaging of biofilm growth in the presence of external flows[21], where measurements using confocal laser scanning microscopy (CLSM) are limited to a penetration depth of less than one hundred microns, whereas OCT can penetrate several millimeters of biofilm. The fast acquisition time of OCT has been used to study the degradation of a biofilm by hydrogen-peroxide over $80\,\mathrm{s}$[18]. Other studies have used time-resolved OCT measurements as a virtual rheometer, measuring the response of biofilm to changes in shear stress to estimate the viscoelastic properties of biofilms[22]. These measurements were later expanded by adding numerical simulations of fluid-structure-interactions[23]. More recently, the modularity of OCT-systems was used to create automated setups capable of monitoring multiple channels over extended periods of time[24,25].

A recent application of OCT has been its use in investigating interfacial flows in millifluidic devices. OCT enables tracking fluid-fluid interfaces as they deform in space and time under shear flow. The D-OCT technique complements this capability by measuring velocity fields in both the streamwise and wall-normal directions within a 2D plane. By combining interface tracking with velocimetry, OCT enables detailed characterization of interfacial phenomena, such as droplet dynamics, interface slippage, and Marangoni flows. Similar to biofilms and non-Newtonian fluids, OCT addresses critical instrumentation gaps for studying interfacial flows by providing high-resolution measurements of mesoscopic systems in dynamic and confined conditions.

We have touched on only a fraction of the investigations where OCT has been applied in soft matter systems. More importantly, OCT has the potential to become an important tool in the community, expanding into new areas of soft matter research. This underpins the motivation for this tutorial review. Our goal is to provide a pedagogical guide focused on the technical aspects of OCT, offering insights into common pitfalls and how to avoid them. Additionally, we showcase a variety of applications to demonstrate how OCT can be employed to study increasingly complex systems. These examples aim to help readers appreciate the breadth of problems that can be tackled using OCT.

This review is organized as follows. Section 2 provides a broad overview of the working principles of OCT and D-OCT. We explain how spatial resolution and temporal resolution are determined by the light source and optical configuration, presenting

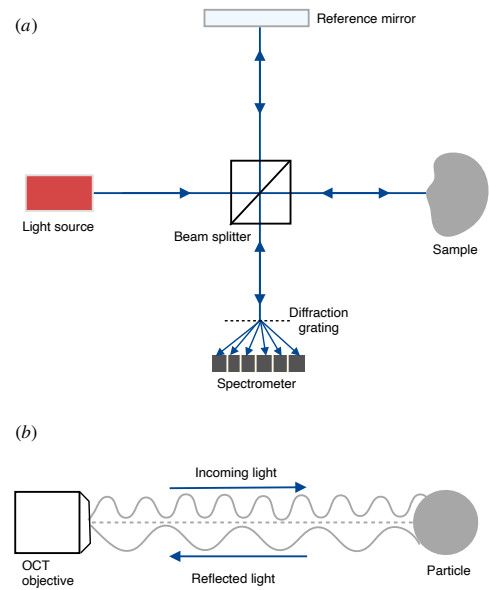

Fig. 1 (a) The basic setup of a Sd-OCT system, where the light beam from the source is split into a reference beam and a sample beam. After reflection, the two beams are recombined before they are split by a diffraction grating and detected by a spectrometer. See Supplementary Figure 1 in the supplementary material for more details. (b) In Doppler-OCT, the frequency shift between the incoming and reflected beam can be used to measure the particle's velocity.

only the essentials needed for understanding the techniques. Section 3 features six tutorials, each supported by Python scripts. These tutorials, summarized in Table 2, cover both intensity- and Doppler-based methods. Section 4 explores a range of applications of OCT relevant to soft matter and fluid mechanics, including time-resolved velocimetry (4.1), particle-based velocity measurements (4.2), slip velocity (4.3), shear-induced structure detection (4.4), fluid-structure interactions (4.5), and fluid-fluid interface studies (4.6). Section 5 offers practical guidelines derived from our experience with OCT, providing a brief but valuable reference for practitioners. The review concludes with a discussion and outlook in Section 6.

## 2 Technical Background

This section outlines the fundamental working principles of OCT for imaging and velocimetry in fluid mechanics and soft matter physics. For detailed exposition of the underlying physics and technical aspects of OCT, we refer the reader to the excellent reviews by Fercher et al.[26] and Tomlins and Wang[11].

### 2.1 Imaging

When a light beam is propagated within a medium, the physical properties of the medium will re-modulate the phase and amplitude of the scattering light[27]. In an OCT apparatus, a beam splitter divides the emitted low-coherence beam coming from a broadband light source into two beams (Fig. 1a). One passes through the sample, while the other is kept as a reference. The two beams are then recombined and construct an interference signal[11]. The output is an intensity signal showing the presence and extent of contrast points inside the medium along the beam.

| Device | Type | $\lambda_0$ [nm] | $\Delta z$ [μm] | $\Delta x$ [μm] | $z_{max}$ [mm] | $f_s$ [kHz] |
| --- | --- | --- | --- | --- | --- | --- |
| Telesto II | SD-OCT | 1300 | 3.4 | 4-12 | 3.5 | 5.5-76 |
| GAN610-SP5 | SD-OCT | 930 | 6.0 | 7-20 | 2.7 | 5.5-248 |

Table 1 Technical specification of OCT devices employed to generate the tutorials and showcases. Axial resolution $\Delta z$ and imaging depth $z_{max}$ are dependent on the refractive index of the medium. The lateral resolution $\Delta x$ depends on the optics mounted to the OCT probe.

A Time-domain OCT (Td-OCT), the first OCT technique developed, uses a scanning mirror to adjust the optical path length difference between the two arms. Through coherence gating, the short coherence length of the light source creates a constructive interference pattern. This pattern appears at mirror positions corresponding to the depth position of the contrast point in the medium[28]. On the other hand, a Spectral-domain OCT (Sd-OCT) operates with a fixed reference mirror. As shown in figure 1(a), in the Sd-OCT, the superimposed back-scattered optical arms (reference and sample) are guided to a 1D-spectrometer through a diffraction grating element[29,30].

Assuming an ideal configuration, the resulting intensity $I(\omega)$ obtained from the spectrometer can be written as

$$I(\omega) = \frac{1}{4}|s(\omega)|^2 (H(\omega)+1)^2. \quad (1)$$

This equation contains the two key ingredients of OCT; i) the spectral distribution of the light source $s(\omega)$ which determines the axial resolution and the penetration depth of the OCT and ii) the frequency response of the sample $H(\omega)$, which is to be reconstructed. The source can be assumed to have a Gaussian distribution and its full width at half maximum (FWHM) can be used to define a coherence length

$$l_c = \frac{4\ln 2}{\pi}\frac{\lambda_0^2}{\Delta\lambda}. \quad (2)$$

Here, $\lambda_0$ and $\Delta\lambda$ are, respectively, the wavelength and optical bandwidth of the source. The length $l_c$ describes the span of a window, over which the low coherence light wave can be approximated as coherent, meaning its phase remains synchronous. The nominal axial resolution (along the OCT beam) in vacuum can be estimated by

$$\Delta z = \frac{l_c}{2} = \frac{2\ln 2}{\pi}\frac{\lambda_0^2}{\Delta\lambda}. \quad (3)$$

If the spectrometer consists of $N$ detectors (e.g. pixels of the CCD sensor), then a discrete Fourier transform of the recorded intensity spectrum results in a time-discrete intensity $I(i\Delta t)$, where $i = 0,1,\ldots,N/2$ [28–30]. In Sd-OCT, the time $i\Delta t$ corresponds to the depth position in the sample. This means that the maximum scanning depth is given by

$$z_{max} = \left(\frac{N}{2}\frac{c}{n}\Delta t\right)\frac{1}{2} = \frac{N}{4n}\frac{\lambda_0^2}{\Delta\lambda}. \quad (4)$$

Here, $c/n$ is the speed of light in a sample material with an average refractive index of $n$. In the above equation, we have approximated $\Delta t = 2\pi/\Delta\Omega$, where $\Delta\Omega$ is the spectral width of the detector $2\pi c\Delta\lambda/\lambda^2$.

An important advantage (e.g. in comparison to confocal microscopy) of OCT is the decoupling of axial and lateral resolutions. The latter is determined by the size of the illumination spot and is influenced by the optics of the OCT probe. The lateral resolution is set by either the numerical aperture of the microscope objective $NA$,

$$\Delta x = 1.22\frac{\lambda}{2NA} \quad (5)$$

or by the beam waist (spot size). For a Gaussian beam we have

$$\Delta x \approx \frac{\lambda \hat{f}}{D}, \quad (6)$$

where $\hat{f}$ is the focal length of the lens, and $D$ denotes the diameter of the incident beam. The scanners used in the object head of current instruments face technical limitations that make it challenging to handle beam diameters exceeding a few millimeters. As a result, lateral resolution is generally limited to ranges lower than the axial resolution. This makes the OCT technique particularly effective for studying stratified objects, where axial resolution plays a key role, or is the main target of probing, in ensuring accurate image interpretation[28]. Note that the maximum depth may be restricted by the $NA$, as the depth of field of an objective is given by

$$z_{max} = 2\frac{\lambda n}{NA^2}. \quad (7)$$

Thus, while high $NA$ provides lateral resolution through (5), it may limit the axial range, if $z_{max}$ in eq. (7) is smaller than the value determined by the light source in eq. (3).

The tutorial examples that follow in section 3 are all obtained from two Sd-OCT devices; Telesto II and Ganymede GAN610-SP5 (Thorlabs, USA). The specifications of these two devices are provided in Table 1. In conventional OCT nomenclature, 1D-probing is termed *A-Scan*. Stacking laterally obtained A-Scans leads to a 2D-recording known as a *B-Scan*. A volume probing, obtained through stacking B-Scans is termed a *C-Scan*. A common way to export and store a series of B-scans or individual volume scans is using the *TIFF* format, specifically *TIFF* image-stacks. These image-stacks contain grayscale images and are highly compatible with a multitude of software packages, such as ImageJ[32], scikit-image[33], or MATLAB[34].

## 2.2 Velocimetry

D-OCT is an extension that enables the measurement of fluid flow velocity profiling by analyzing the frequency shifts in the OCT signal caused by the Doppler effect (Fig. 1b).

The classical approximation of the Doppler effect (where relativistic corrections are negligible) can be written as a shift in

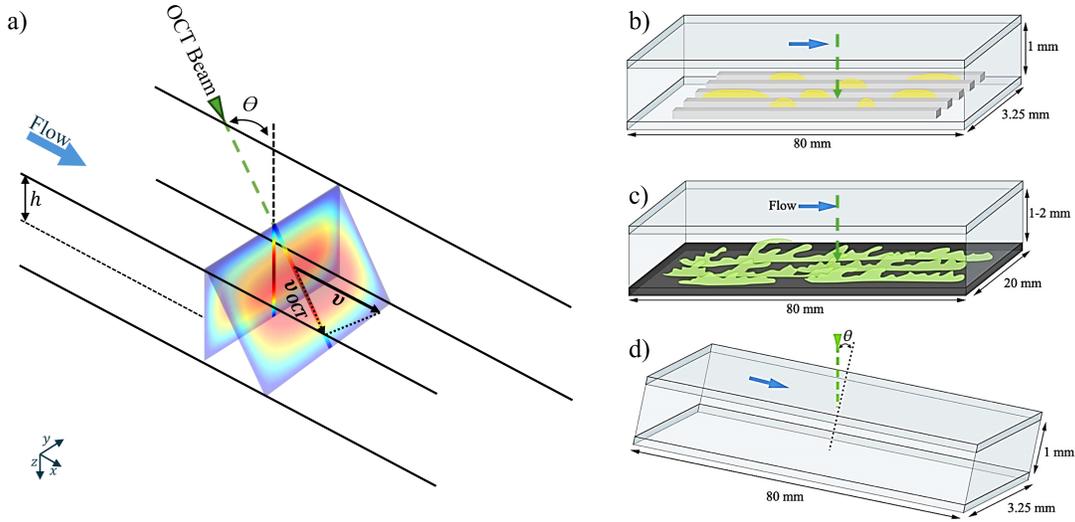

Fig. 2 Schematics of the flow configurations used in the tutorials. a) Inclination of the beam with respect to the local flow direction reconstructed from Amini et al.[31]. b) Wall with grooves infused with a second fluid phase. c) Millifluidic system with a biofilm growing on the wall. d) Rectangular duct inclined for Doppler-based velocimetry.

frequency

$$\Delta f = \frac{2v \sin \theta}{\lambda_0}, \quad (8)$$

where $\theta$ measures the inclination of the channel (See Figure 2). The OCT, however, measures the phase shift, which can be obtained from eq. (8)

$$\Delta \Phi = 4\pi \frac{v \sin \theta}{\lambda_0 f}, \quad (9)$$

where $f$ is the acquisition frequency. It is evident that for the Doppler measurement to capture a non-zero component of the flow velocity, the emitted beam and the local velocity vectors should not be orthogonal. As illustrated in Figure 2(a), there exists a compromise between a higher contribution of the flow velocity along the OCT beam, in large inclination angles, and the precision in positioning of the point, over which the velocity profile is obtained. Therefore, by incorporating the inclination and the refractive index of the medium, the following can be written for the velocity in the longitudinal direction,

$$v = \frac{\lambda_0 f \Delta \Phi}{4n\pi \sin \theta}. \quad (10)$$

From the expression above, we may estimate the velocity to $v \sim 1$ cm/s, since $\lambda_0 \sim 10^3$ nm, $f \sim 100$ kHz, and $\Delta \Phi / 4n\pi \sin \theta \sim 1$.

## 3 Tutorials

This section presents tutorials on the intensity- and Doppler-based methods. It is assumed that an OCT scan has been successfully acquired and that an intensity- or phase-field has been saved in the form of a *.tiff* image-stack. The tutorials describe how the data may be processed to derive physically meaningful quantities for a given application. Table 2 summarizes the tutorials presented in this section.

### 3.1 Intensity-based Schemes

#### 3.1.1 Interface Detection (high contrast)

In the simplest configuration, OCT can be used to detect a clearly defined interface, for example, between a liquid and a solid surface or between two immiscible liquids. Here, we detect the shape of a submerged oil droplet in milk diluted in water (20%). In this first tutorial, there is a clear separation of the signal into foreground and background. Figure 2(b) shows the configuration, where the bottom wall of a millifluidic channel is structured with longitudinal grooves with cross-section of $243 \times 376 \, \mu m^2$. The grooves are infused with oil (hexadecane), and a solution of water and milk is flowing over them at a constant flow rate of $0.4 \, \text{mL min}^{-1}$.

Capillary instabilities break the water-oil interface into droplets with a length of a few millimetres. Due to interface pinning, the droplets remain trapped in the grooves when exposed to flow. The contrast between water and lubricant phases is created by milk colloids, which are only miscible in water. These micro-sized particles act as a seeding medium, scattering the infrared light of the OCT beam. Consequently, the milk solution appears bright (due to scattering), while the transparent fluid appears dark (no scattering), as in Figure 3(a). Other solutions, such as water-soluble food colouring or coffee, can also render the fluid opaque. In a fluid flow, selecting a contrast agent that is not surface-active is important to avoid Marangoni flow which could potentially modify the interfacial dynamics.

To obtain the interface shape, the flow cell is positioned under the OCT probe at a working distance corresponding to the focal length of the optics. Note that common materials for the optical access window, e.g. PMMA or glass, have high refractive indices, effectively increasing the distance between the probe and the target. Flat solid-liquid or solid-solid interfaces can reflect the beam vertically toward the sensor, causing autocorrelation noise. Therefore it is beneficial to incline the sample. An angle of approximately 6° between the beam and the plane of the channel is sufficient to deviate the reflections.

| | Tutorial | Application | Code |
|---|---|---|---|
| 1 | Interface detection (high contrast) | Oil droplet immersed in a water-milk mixture | t1_simple_thresholding.py |
| 2 | Interface detection (low contrast) | Biofilm growth in flow | t2_complex_thresholding.py |
| 3 | Wall detection in opaque media | Water-milk mixture between two solid (PMMA) surfaces | t3_t5_DOCT_1D.py |
| 4 | Particle detection in transparent media | Rigid particles in a dilute polymeric fluid | t4_particle_detection.py |
| 5 | A-Scans: 1D Velocimetry | Laminar duct flow | t3_t5_DOCT_1D.py |
| 6 | B-Scans: 2D Velocimetry | Laminar duct flow | t6_DOCT_2D.py |

Table 2 List of six tutorials presented in section 3. The problem considered by each tutorial is also listed alongside the name of the associated Python script.

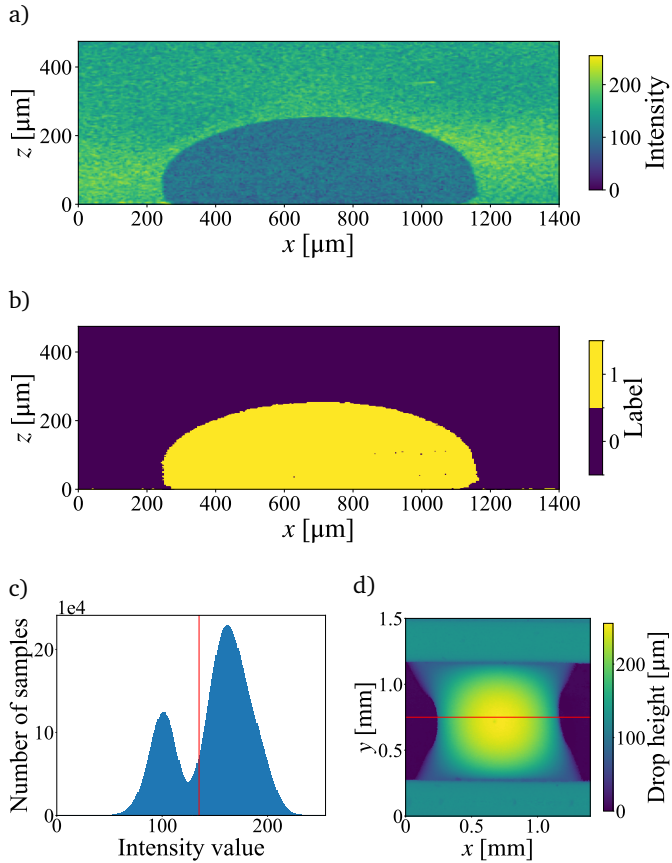

Fig. 3 *Tutorial 1:* An oil droplet in exposed to fluid (20% milk, 80% water) flow in a channel. a) Raw intensity field along the centerline of the channel. b) Isolated droplet after processing. c) Intensity histogram of the full measurement with binarization threshold in red. d) Resulting height map of the droplet surface confined between two ridges, visible in the upper and lower part. The red line marks the location of the slice in a) and b).

A full 3D scan (C-scan) is obtained through a series of B-scans (two-dimensional planes) within the field of view. The measurement is extracted from the proprietary .oct format and saved as a *.tiff* image stack. The image analysis protocol implemented in *t1_simple_thresholding.py* demonstrates how to separate the profile of the drop in the foreground from the background, to obtain Figure 3(b). First, the image stack is loaded and trimmed to the relevant region of interest. The milk-oil-interface is characterized by a significant drop in the signal intensity (Fig. 3(a)). This results in a clear bimodal intensity distribution (Fig. 3(c)), which lends itself to classical image binarization techniques, such as Otsu's method[35]. With this approach, the optimal threshold is determined by minimizing intra-class variance, or equivalently, maximizing inter-class variance between the two pixel intensity distributions. Otsu's algorithm computes the image histogram and iterates over all possible threshold values to identify the one that best separates the pixel classes.

The resulting threshold is shown as a red line in Figure 3(c). Due to the noisy nature of OCT measurements, salt-and-pepper noise will likely cause small spurious signals to be present. These are removed based on their size, and the droplet is selected as the largest structure within the scan (Fig. 3(b)). From this, a height map depicting the interface is calculated (Fig. 3(d)). This tutorial thus demonstrates how OCT can be applied to obtain spatial information of interfaces in 3D complex geometries. OCT is particularly convenient when there is a strong reflectivity at the interface. This results in a bimodal intensity distribution, where classical binarization techniques can be used.

### 3.1.2 Interface Detection (low contrast)

In many OCT scans, traditional thresholding approaches fail to separate foreground and background. If the two signals are separated by only a small difference in intensity compared to the noise level, the intensity distribution will not be bimodal, and therefore not suited for e.g. Otsu's method. In this second tutorial, we show the use of OCT for in-situ measurements of a bacterial biofilm in a fluid flow as an example of such a signal (See *t2_complex_thresholding.py*). Biofilms consist of cells that are embedded in a matrix of extracellular polymeric substances[36]. These cells attach to an interface, here the wall of a laminar flow cell, and grow into complex structures. Since the individual cells are smaller than the resolution of the OCT, the biofilm can be treated as a continuous material. In this example, we separate the two phases, i.e. the biofilm and the surrounding medium, using a thresholding algorithm that is informed by the characteristics of the OCT scan. The resulting intensity field can be used to calculate the characteristics of the biofilm, such as the volume.

Figure 2(c) shows the configuration, where biofilm develops on the bottom wall of a channel in presence of liquid flow (flow rate 60 mL min$^{-1}$). A 2D vertical slice of a volume scan containing biofilm is shown in Figure 4(a), which can be processed to yield the isolated biofilm (Fig. 4(b)). The OCT beam is traversed over the field of view by two mirror galvanometers. Since these mirrors cannot occupy the same physical space, the actual path length of the beam within the OCT device depends on the current position of the mirrors. This causes, especially for large fields of view, a local offset in the signal, warping the acquired data. Thus, the substratum must be detected and aligned with the bottom of the scan to yield correct height information. The scan is characterized by a small amount of signal compared to the background. Therefore, no second peak can form in the intensity histogram (Fig. 4(c)), and Otsu's method is not able to correctly separate the biofilm from the background.

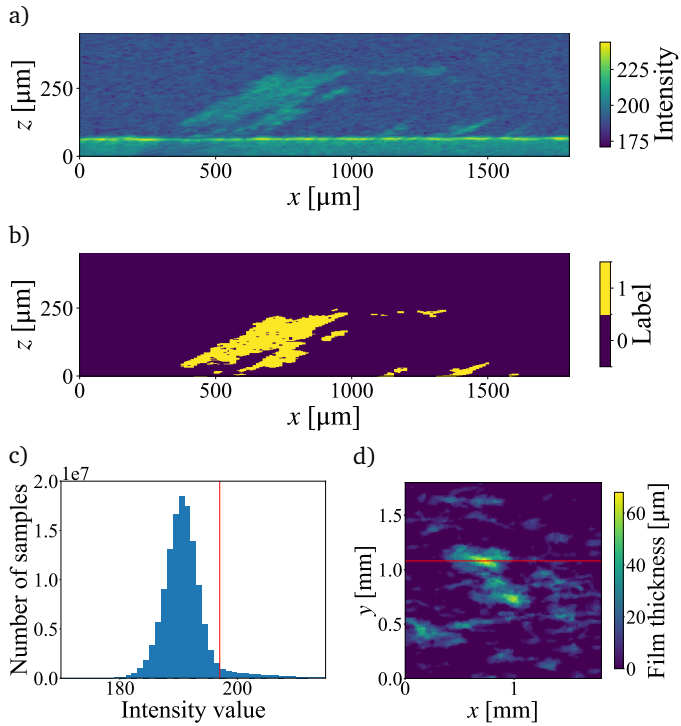

Fig. 4 *Tutorial 2:* Biofilm growing on a flat surface in a shear flow. a) Raw intensity field along the red line in d). b) Isolated biofilm after processing. c) Intensity histogram of the full measurement with binarization threshold in red. d) Resulting height map of the droplet surface. The red line marks the location of the slice in a) and b).

Alternatively, a filter can be implemented based on the known shape of a typical scan containing small samples. The inherent noise in an OCT scan containing only background signal will create a Gaussian intensity distribution. Here, the foreground has a higher intensity, creating an inflection point in the intensity histogram that can be used to detect the threshold. The resulting threshold is again shown as a red line in Figure 4(c). Finally, small, mostly floating, structures are eliminated via a modified median filter, that does not fill in any gaps. This filter compares the value of each pixel with its neighborhood and changes it from one to zero, if the median value within this neighborhood is zero. The resulting dataset contains the full three-dimensional geometry of a biofilm, enabling the calculation of parameters such as biofilm thickness (Fig. 4(d)), total biovolume, roughness metrics, or porosity.

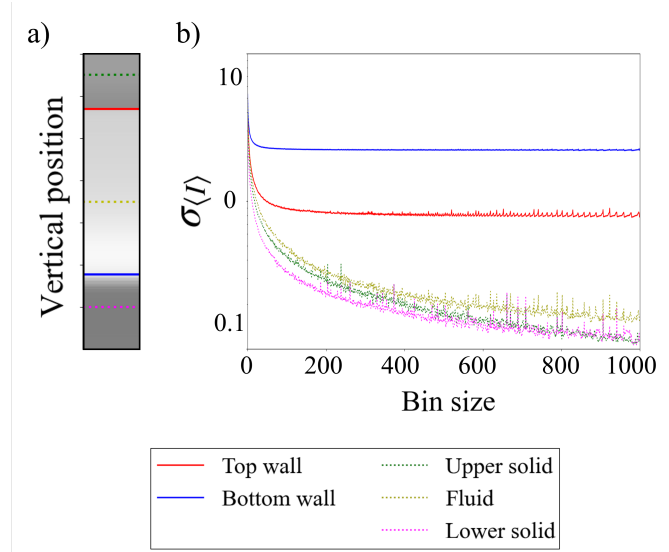

Fig. 5 *Tutorial 3:* Wall detection in a milk-filled channel. a) Mean intensity along an A-scan. b) Standard deviation of the mean intensity depending on the size of the averaging window, corresponding to the locations in (a).

Since the OCT beam is able to penetrate more than a millimeter into the sample, thick, complex geometries can be resolved, whereas CLSM measurements struggle to penetrate more than a few hundred micrometers of material[37] (see Fig. 2(b)). Additionally, voids within or beneath the biofilm can be resolved, compared to confocal laser scanning microscopy, where apparent voids can also be caused by signal attenuation or limited penetration of the fluorescent dye[37].

### 3.1.3 Wall Detection in Opaque Media

Many applications involve measurements near solid surfaces. In such cases, the location of the wall often needs to be determined with precision. Figure 5(a) shows the vertical distribution of the mean intensity signal obtained when performing an A-scan of a water-milk mixture between two solid (PMMA) surfaces. The upper surface is clear and the lower surface is opaque, to avoid reflections. We observe a diffuse interface between the bottom wall and the liquid, which makes wall detection challenging. In particular, the scattering intensity of the milk is much larger than that of the clear material at the top as well as the opaque bottom of the channel. The interface can be detected via the maximum change of intensity between adjacent pixels. A different – and more robust – approach is to first down-sample the A-scan replicates, then calculate the standard deviation of the resulting intensities in time.

Figure 5(b) shows the standard deviation of the intensity signal $\sigma_{\langle I \rangle}$ for different window sizes used in the down-sampling. The colors of the curves in Figure 5(b) correspond to the vertical locations marked in Figure 5(a). We observe that the signal fluctuations in the bulk of the fluid and the solid (dotted lines) are much smaller compared to the fluctuations at the solid-fluid interfaces (red and blue lines). In addition, the fluctuations become independent of the window size for the positions corresponding to the top (red) and bottom (blue) walls. Thus, one way to determine the locations of the walls is by identifying the local maxima of the standard deviation of the intensity signal obtained from an A-scan.

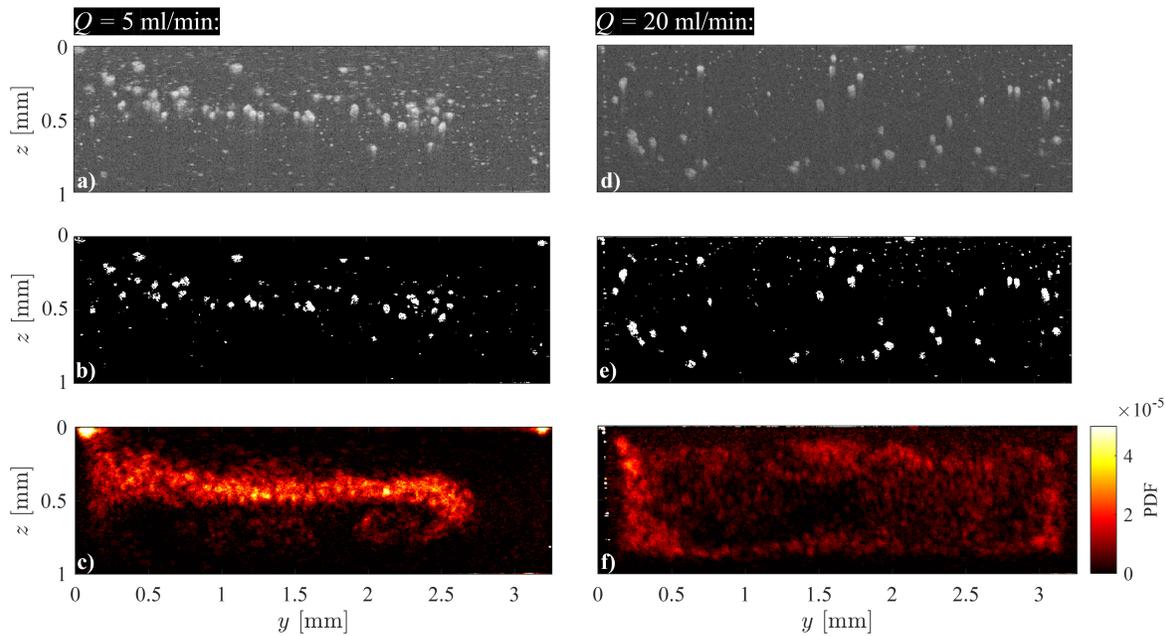

Fig. 6 *Tutorial 4:* Polyacrylamide (PAA) 210 ppm, with 0.8 % wt/vol 56 µm PSP polyamide tracer particles. The flow is directed into the plane. a,d) Sample raw intensity field. b,e) Binarized frame. c,f) Probability distribution function of the particles over 500 frames. a,b,c) $Q = 5\,\text{mL}\,\text{min}^{-1}$, d,e,f) $Q = 20\,\text{mL}\,\text{min}^{-1}$ [38].

### 3.1.4 Particle Detection in Transparent Media

The intensity field obtained by OCT can be used to detect opaque particles dispersed in a transparent fluid. This enables investigating dynamics of non-colloidal particles migrating with relative velocities with respect to the fluid flows or using OCT as a particle-based flow measurement technique (e.g. PIV, PTV, etc.) by imaging the time-dependent dynamics of the *faithful* tracer particles. The former application is discussed in this section (*t4_particle_detection.py*) while an example of the latter is presented in section 4.2.

The specific tomographic field of view accessible to OCT provides complementary information to the conventionally-used, top-view fluorescent microscopy in microfluidics[39]. Figure 6 shows OCT intensity fields that were obtained when investigating the migration of rigid particles in a dilute polymeric fluid[38]. Measurements were taken at a streamwise position of 60 mm, in a rectangular duct (cross-section $1 \times 3.25\,\text{mm}^2$). The long-chain polymeric solution (Polyacrylamide, PAA) – transparent at the given concentration of 210 ppm – is detected as a weak source of contrast. In the raw intensity images shown in Figure 6(a) and (d), traces of the polymeric network is visible as small dots. The more prominent bright spots correspond to PSP polyamide tracer particles (56 µm in diameter and dispersed at a volume fraction of 0.8 % wt/vol). Two observations can be made. First, opaque particles have a better-defined top surface, as the OCT beam decays as it penetrates the solid, opaque particle. For relatively small particles, this effect will be negligible. Second, these opaque particles will cast a shadow within which the back-scattered signal will be dimmer. This will impose an upper limit on the volume fraction of the opaque particles dispersed in the fluid, as the statistics of the particles in lower-depth positions will be overshadowed by the upper layers.

Figures 6(b) and (e) show the results of image processing applied to the raw intensity fields, while Figures 6(c) and (f) display the probability distribution function (PDF) of particles passing through the cross-section over 500 frames. We observe stationary contrast points, such as particles stuck to the duct walls, bubbles, glare points, and diffraction from solid boundaries. These points affect the scaling and statistics of the calculated PDF.

To address this issue, the first processing step involves subtracting the mean image from each frame in the series. This step was intentionally skipped for Figure 6 to highlight the influence of stationary features. The second step normalizes the grayscale images, adjusting the gray value histogram of each frame to the full span from 0 to 255. Finally, a high-pass filter can sharpen particle interfaces, which is particularly useful for particle-based velocimetry techniques. However, this filtering was also omitted in the preparation of Figures 6(b) and (e).

Next, a threshold is applied to binarize the filtered grayscale images into black-and-white ones. The threshold is determined using the method described in Section 3.1.2. The goal of these preprocessing steps is to retain as many particles of interest as possible while eliminating extraneous optical information and noise captured by the OCT system.

The left column of Figure 6 (with flow rate Q=5 ml/min) illustrate the focusing effect of elasticity at low particle inertia. In the right column, at a higher flow rate (Q=20 ml/min), we ob-serve inertial migration of the particles to a preferential location between the wall and the centerline (Ségre-Silfverberg effect).

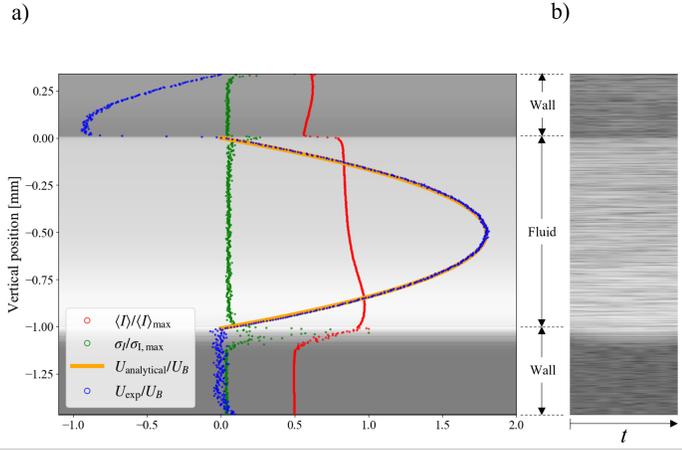

Fig. 7 *Tutorial 5:* Doppler OCT velocimetry. a) Validation of the full duct height velocity profile measured for a Newtonian fluid with the Boussinesq approximation series, and b) raw intensity A-Scan sequence.

### 3.2 Doppler-based Schemes

#### 3.2.1 A-Scans: 1D Velocimetry

Using the Doppler phase shift mode introduced in section 2.2, a velocity profile can be measured along the beam direction. Using the relation between the Doppler phase shift and the velocity in equation 10, we calculate the streamwise velocity profile. Figure 7(a) shows the velocity profile of laminar Newtonian flow in a duct (Fig. 2(d)). Here, the same setup as in section 3.1.3 is used, with a milk-water solution flowing between two PMMA walls. The profile is validated with the analytical Boussinesq solution for rectangular ducts (see Amini et al.[31] for details). The corresponding code can be found in *t3_t5_DOCT_1D.py*.

Figure 7(b) shows the time sequence of A-Scans, in which the top and bottom walls are observable. As shown in Figure 7(a), the local peaks in the intensity signal have been used to detect the exact position of the walls. Here, the depth resolution is $2.58\,\mu m$ along the beam. The duct was positioned with an inclination of $6°$. The profile is obtained for the spanwise middle point of the duct to minimize the effects of the sidewall. The height of the duct is 1 mm, and the volumetric flow rate is $0.4\,\mathrm{mL\,min^{-1}}$.

#### 3.2.2 B-Scans: 2D Velocimetry

Stacking lateral (or longitudinal) 1D A-Scans will result in a 2D measurement of the flow field. The current state of the art allows for A-Scan acquisition frequencies of $\mathcal{O}(10-100)\,\mathrm{kHz}$. The effective frequency (inverse of the time interval between two consecutive recordings) of B-Scans is the A-Scan frequency divided by the number of lateral pixels in the B-Scan plus a small overhead for the fly-back of the mirror galvanometers.

It is important to note that Doppler OCT is inherently event-based, capturing physical information only at instances and locations, where contrast points pass through the test section. While it provides highly accurate instantaneous measurements, these are based on a non-uniform spatial and temporal grid. As a result, the entire velocity field must be corrected using either the volumetric flow rate or a known non-zero velocity at a single point. Gowda et al.[12] used the following formulation for the velocity correction based on the volumetric flow rate

$$U_{\text{corr}}(y,z) = U_{\text{meas}}(y,z) \left( \frac{Q}{\int U_{\text{meas}}(y,z)\,\mathrm{d}y\,\mathrm{d}z} \right), \qquad (11)$$

where $Q$ is the volumetric flow rates set by a volumetric pump or measured by an independent device.

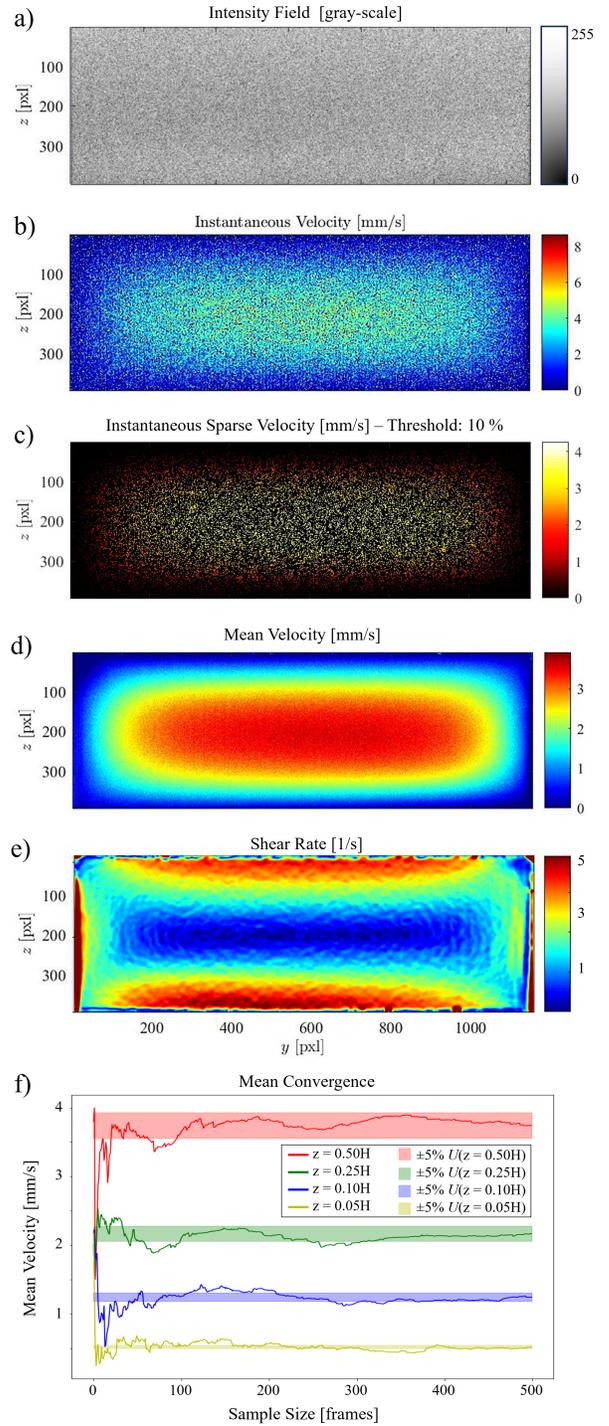

Fig. 8 *Tutorial 6:* Doppler OCT 2D B-Scans. a) Intensity field. b) Instantaneous velocity field. c) Sparse data points within $10\%$ margin of the mean velocity. d) Mean velocity field over 500 frames. e) Shear rate field. f) Convergence of the mean velocity at different channel heights ($5\%$, $10\%$, $25\%$, and $50\%$ of the full height). The shaded regions represent a $5\%$ deviation from the final mean.

Here, we used the setup shown in Figure 2(d). Undiluted milk is used as the scattering medium and the flow cell is a rectangular duct with cross-section $1 \times 3.25\,\text{mm}^2$. Figure 8 shows an example of a 2D velocity field obtained by a D-OCT B-Scan scheme. The associated script is available in *t6_DOCT_2D.py*. All of the following measurements have been trimmed to the interior of the channel. Figure 8(a) shows the raw intensity field, which is captured simultaneously with the Doppler phase shift measurement. Figure 8(b) displays a single-frame instantaneous velocity field, where the scattered nature of the raw data is evident. By limiting the data points to those deviating within a 10% margin from the mean of 500 frames, the sparsity of the raw data becomes apparent (Fig. 8(c)). Finally, Figure 8(d) presents the temporal mean velocity field calculated over 500 frames (i.e., B-Scan recordings). To demonstrate the usefulness of OCT for near-wall measurements, we can compute the mean magnitude of the shear rate field from the mean flow $u$, using

$$\dot{\gamma}(y,z) = \sqrt{\left(\frac{\partial u}{\partial y}\right)^2 + \left(\frac{\partial u}{\partial z}\right)^2}. \qquad (12)$$

This is shown in Figure 8(e). To reduce noise in the signal, a Gaussian filter was applied prior to calculating the gradient. The convergence of the mean velocity at different channel heights along the spanwise centerline is shown in Figure 8(f). Here, it can be seen that in this case, approximately 250 B-scans are required to obtain a converged result. The scattering properties of the medium and the density of contrast agents are key factors in this convergence. Note that convergence rates near walls can be different compared to far-field regions.

## 4 Applications

In this section, we provide several applications in fluid mechanics and soft matter physics, where OCT has been used. We hope that these showcases can provide further insight into the capacity of OCT, but also to inspire its use in problems that are traditionally addressed using other techniques.

### 4.1 Time-Resolved Velocimetry

The acquisition frequency of the OCT is relatively high ($\mathcal{O}(10-100)\,\text{kHz}$) compared to the typical macroscopic time scales in many soft matter systems. Therefore, a wide range of filtering and signal processing methods are applicable to curate the raw, noisy data. The noise is mostly due the sparsity of the event-based measurement, as the contrast agents pass through the test section in a non-continuous manner. In this section, an example of *ad hoc* signal-processing methods to reduce noise is presented for the case of a complex fluid flow susceptible to shear-banding driven instabilities.

In general, as a common flow heterogeneity observed in some non-Newtonian fluid flows, *shear-banding*[40] is related to meso-structural re-configuration the material undergoes in response to local shear. It has been shown that shear-banding fluids are prone to instabilities[41,42]. Here, a pNIPAM-based microgel is studied with Time-Resolved OCT-based velocimetry (hereafter TR-DOCT). The sample fluid shows a high yield stress and weak footprints of shear-banding[43] in its rheological flow curve. It should also be noted that the opaqueness of such fluids with packed internal structures (i.e., a 3D network of cross-linked microgel particles) would generally not allow the use of conventional velocimetry techniques (e.g., PIV, PTV, LDV, etc.). However, OCT measurements are possible without any added contrast agents.

Figure 9(a) (red curve) shows the raw time signal of the normalized fluctuation velocity for a sample point located at the half-height of a duct (see Figure 2) over 9800 recordings at 5.5 kHz. The raw signal shows sporadic unphysical large peaks. A scheme based on the second derivative of the signal can be used to detect the outliers (Fig. 9(b)). The outliers, identified by the peaks in the second derivative signal, are replaced through interpolation of their neighboring values. To determine the optimal threshold for detecting these outliers, a sweep of the threshold values is performed while monitoring variations in the *rms* of the resulting signal. It was observed that the control parameter $u_{rms}$ shows two distinct drops as the threshold is reduced. The first drop corresponds to the removal of high-frequency noise (spurious outliers), while the second drop is associated with filtering out the actual physical portion of the time-resolved signal. The plateau between drops can be used to estimate the optimal threshold, balancing effective noise removal and signal preservation. The resulting outlier-free signal is shown in green color throughout Figure 9. It should be noted, that this spectral decoupling is due to high frequency ranges of the instrument noise, compared to the relatively low frequency of the physical fluctuations in this millifluidic channel flow at low inertia.

Furthermore, through *a priori* knowledge of the noise spectrum, *ad hoc* tailored filters can be implemented to the outlier-free signal. Here, we have implemented a low-pass filter with a cut-off frequency of 50 Hz to isolate fluctuations from large scale structures passing through the test section. The filter was a 14th-order Infinite Impulse Response (IIR) design, with a passband frequency of 50 Hz, stopband frequency of 65 Hz, passband ripple of 1 dB, and stopband attenuation of 30 dB. Figure 9(c) shows the original signal in red and the velocity signal after outlier removal and low pass filtering in black. Figure 9(d) shows the histogram of the raw (red), outlier-free (green) and large-scale low pass filtered signals (black). Implementing the above steps on the whole half-height velocity profile, Figure 9(e),(f) and (g) illustrate the raw, outlier-free and low pass filtered signals, respectively.

### 4.2 Particle-based Velocimetry using OCT

Using high frequency B-Scans in a vertical plane parallel to the flow direction, the intensity field obtained from OCT can be used for 2D2C (i.e., 2 velocity components on a 2D field of view) particle-based velocimetry schemes such as planar PIV and PTV. Unlike section 3.1.4, where macroscopic particles with non-zero

Journal Name, [year], [**vol.**], 1–17 | 9

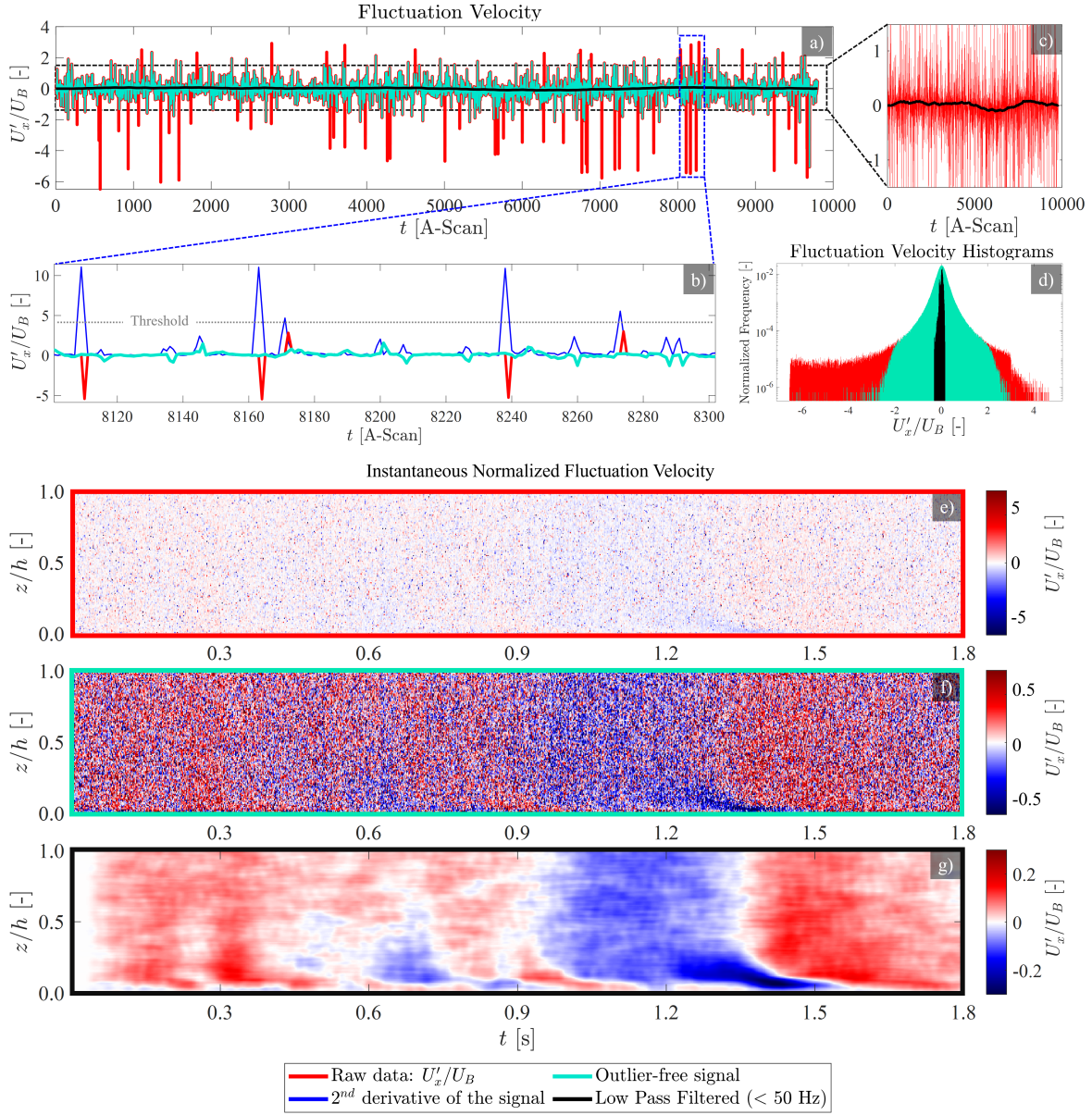

Fig. 9 Time-resolved Doppler velocimetry. a) Raw time-resolved signal for a single point located at the half-height elevation of the duct, over 9800 consecutive A-Scans at 5.5 kHz. b) The peaks in the second derivative of the signal have been used as indicators of the outliers in the signal. c) Low pass filtered signal, allowing frequencies below 50 Hz. d) Histogram of the raw, outlier-free, and low pass filtered signals. e) Spatiotemporal presentation of the raw time-resolved velocity profiles over half-height of the duct for 9800 A-scans (equivalent to 1.8 s). The sporadically distributed outliers range over $\pm 500\%$ of the bulk velocity $U_B$. f) Outlier-free spatiotemporal presentation of the velocity profiles, limiting the velocity variations to $\pm 50\%$ of the bulk velocity $U_B$, and g) using a low pass filter along time axis, and a moving average along the spatial axis to extract large scale structures of the fluctuation velocity.

relative velocities compared to their surrounding fluid medium were considered, here small, neutrally buoyant tracer particles are used, which faithfully follow the flow.

More specifically, 220 ppm of Titanium Dioxide (TiO$_2$) particles (nominal diameter 1 μm) were dissolved in DI-water. Figure 10 shows the particle-based velocimetry in a $1 \times 1 \, \text{mm}^2$ duct with a 4 mm-long rectangular cavity of $1 \times 1 \, \text{mm}^2$ cross-section. The cavity is located 40 mm from both the inlet and the outlet of the duct. A constant flow rate of $0.2 \, \text{mL} \, \text{min}^{-1}$ is maintained and the OCT intensity A-Scans are stacked in a vertical plane along the flow di-rection in the mid-span of the duct and the cavity, making a 1000 pixel wide B-Scan. Moreover, the measurement were performed at 76 kHz A-Scan acquisition frequency for 500 consecutive frames (B-Scans) spanning over a duration of 6.58 s. Pixel dimensions are 2.58 μm in depth and 5 μm in the longitudinal directions. In order to guide the reflections away from the receiver, the setup is tilted at a small angle.

We consider a 2 mm depth of the field of interest. The PIV/PTV input frames are obtained by superposing five series of measurements acquired with the OCT focal plane sweeping in the depth direction (Fig. 10(a)). The superposition of the frames, shown in Figure 10(b), assumes that the flow field is steady. A snapshot of the mean-subtracted stacked intensity field (Fig. 10(c)) is then normalized to fit the grayscale range [0,255] (Fig. 10(d)). A high-pass filter (HPF) is then applied to the image to sharpen the particle interfaces (Fig. 10(e)), before the image is binarized (Fig. 10(f)). Given the steady flow field, 5 frames (each 5 frames apart) are superimposed to increase the tracer particles density (Fig. 10(g)).

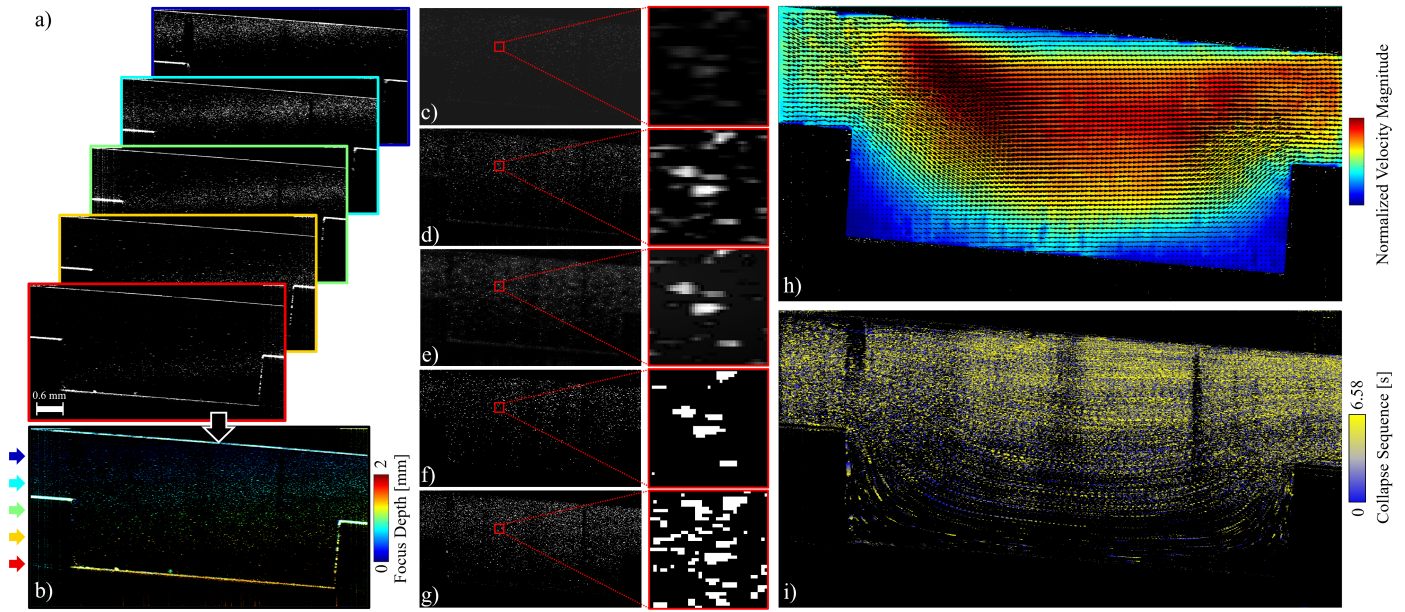

Fig. 10 Particle-based velocimetry schemes based on OCT intensity B-Scans in flow direction on the mid-span plane of a cavity. a) Snapshots of the intensity measurements at different focal plane depths. Frame colors correspond to the color bar in (b). b) Superposition of the snapshots in (a), color coded based on the depth of the focal plane. c) Sample gray-scale input frame, after subtraction of the mean (over 40 frames). d) Normalized gray-scale image to the range [0,255]. e) High pass filtered image, with sharpened particle edges. f) Binarized image (black-and-white scale). g) Superposition of 5 frames, with intervals of 5 frames for independent sampling. h) Particle Image Velocimetry (PIV) results for 2D2C velocity field. i) Lagrangian Particle Tracking (LPT) results for trajectories of the particles obtained through collapsing a sequence of 40 frames.

Figure 10(h) shows the mean flow field obtained through post processing the images with a PIV script. Three passes have been used in searching for correlation of interrogation windows (IWs) of 64×64, 32×32, and 16×16 pxl$^2$, respectively. The search windows (SWs) are set to 2 times the dimensions of the IWs at each pass. The steps (i.e., increments between the center point of neighbor IWs) are set to 50% overlap. The same images were also used for Lagrangian Particle Tracking (LPT) through collapsing a time sequence of 40 processed images (Fig. 10(g)), shown in Figure 10(i) as a proof of concept.

### 4.3 Slip Velocity

Slip refers to the phenomenon where the velocity of a fluid differs from that of the solid surface in contact with it[44]. Depending on the length scale considered, slip can be categorized as either intrinsic or apparent. Intrinsic slip occurs at the molecular level, where fluid constituents are effectively sliding on the solid surface. Experimental measurements have reported molecular slip lengths of the order of tens of nanometers[45–52]. Commonly used measurement techniques are colloidal probe atomic force microscopy (AFM)[53,54] and surface force apparatus (SFA)[46,49], which provide nanometric resolution. Apparent slip, in contrast, is typically observed at larger scales, such as in flows over gas or liquid layers. The slip lengths measured using conventional techniques typically range from 1 to 100 $\mu$m[55–59]. With a spatial resolution of $\mathcal{O}(1)\mu$m, OCT is well suited to measure apparent slip. In the following, we will refer to apparent slip simply as *slip*.

#### 4.3.1 Single Phase Yield Stress Fluids

Yield-stress fluids (for example, microgels and concentrated emulsions) tend to slip on smooth surfaces, and an unknown degree of slip disturbs both rheological characterisation of yield-stress fluids and flow measurements. Figure 11(a) shows an example of a 2D B-Scan on the cross-section of a rectangular duct. A Yield Stress Fluid (YSF) has been used, in which the internal structure of the complex fluid provides the required opaqueness for the OCT measurements, i.e., no contrast agent (dyes or particles) needs to be added. The stress versus strain behavior of this class of complex fluids can be modeled with the Herschel-Bulkley constitutive equations:

$$\dot{\gamma} = \begin{cases} 0, & \tau \leq \tau_y \\ \sqrt[n]{\dfrac{\tau - \tau_y}{\kappa}}, & \tau > \tau_y. \end{cases} \qquad (13)$$

Here, $\tau_y$ denotes the yield stress (i.e., the thresholding limit of stress required to fluidize the material), while $\kappa$ and $n$ represent, respectively, the consistency parameter and the power-law exponent of the model[60]. Figure 8(b) compares the velocity profile measured below the top wall (see the dotted line in Figure 8(a)) with the analytical solution obtained by inserting the Herschel-Bulkley model (13) into the stress term in the Navier-Stokes equations[61]:

$$u(z) = \begin{cases} u_s + \dfrac{nG^{1/n}}{(n+1)\kappa^{1/n}}\left[z_y^{\frac{n+1}{n}} - (z_y-z)^{\frac{n+1}{n}}\right], & 0 \leq z \leq z_y \\ u_s + \dfrac{nG^{1/n}}{(n+1)\kappa^{1/n}} z_y^{\frac{n+1}{n}}, & z_y \leq z \end{cases} \quad (14)$$

where $G$ is the streamwise pressure gradient and $\kappa$ and $n$ denote the Herschel-Bulkley constitutive model fit parameters. Here, $u_s$ represents the slip velocity, as a prominent slippage on the wall is observed in YSFs, and the yielding height $z_y$ (i.e., the elevation of the starting point of the unyielded plug region in the center of the duct, where the local shear stress is below that of the yielding limit) is obtained by:

$$z_y = \frac{\tau_w - \tau_y}{G}. \quad (15)$$

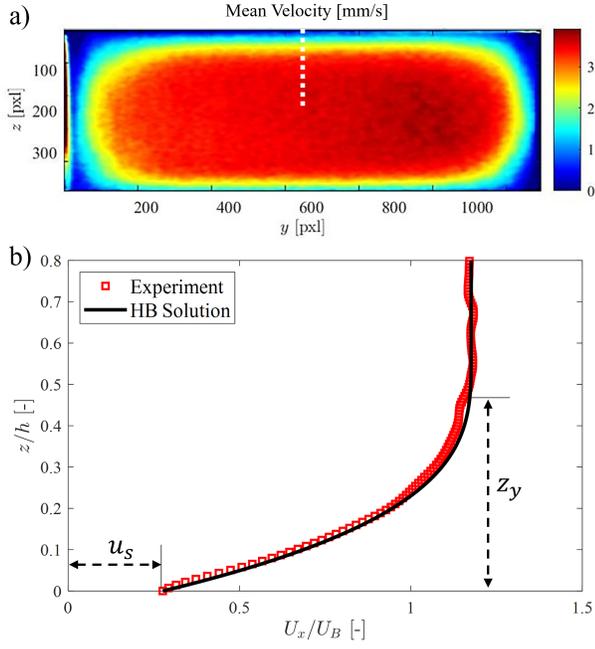

Fig. 11 A sample YSF in a rectangular duct. a) Cross-sectional velocity field. b) Validation of the velocity profile obtained on the mid-span of the duct with the analytical solution Eq. (14) based on the Herschel-Bulkley constitutive equation.

### 4.3.2 Fluid Interfaces

In addition to non-Newtonian fluids, slip occurs in situations where there is a fluid-fluid interface. Examples include superhydrophobic surfaces (SHSs) – where the interface is between water and air – and lubricant-infused surfaces (LISs), where water flows over an immiscible liquid, often an oil. With OCT, near-interface velocity and interface shape can be measured simultaneously, allowing for accurate local slip length calculations. Figure 12 illustrates slip length measurement over a LIS in a rectangular duct. The rectangular grooves are infused with hexadecane and aligned parallel to the overlying water flow. The grooves have a width of $223\,\mu m$, a depth of $237\,\mu m$, and are $80\,mm$ long, corresponding to the length of the rectangular duct. The cross-section of the duct is $1 \times 3.25\,mm^2$.

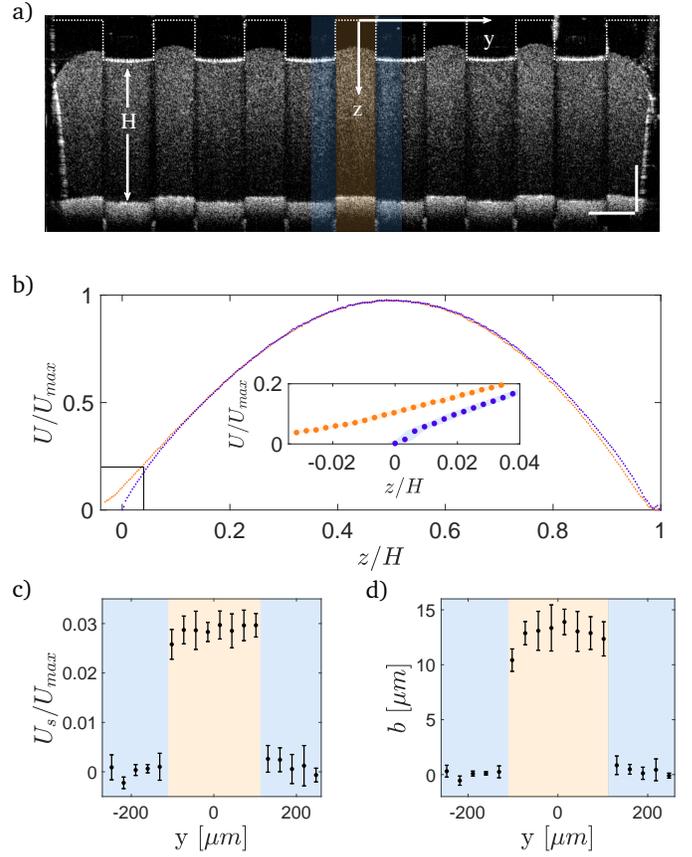

Fig. 12 Experimental measurement of slip length on LIS. a) Intensity image of the cross-section of the rectangular duct. At the upper wall the oil infusing the grooves appears black and water appears gray. The colored areas correspond to the probed regions. The scale bar equals $250\,\mu m$. b) Exemplary velocity profiles on a water-solid area (blue) and on a water-oil area (orange). The colored intervals around the markers represent the standard deviation of the measurements. c) Slip velocities normalized with the maximum velocity with respect to the spanwise position. d) Local slip lengths with respect to the spanwise position.

Water is mixed with bovine milk (volume ratio of 5:1) to distinguish it from the lubricant phase (Fig. 12(a)). Additionally, milk provides the scattering particles needed for Doppler measurements. A series of A-scans were performed in the spanwise direction, from one central ridge to an adjacent one, covering the water-lubricant, and water-solid interfaces. Figure 12(b) shows two profiles; red curve corresponds to profile at water-solid boundary, where the velocity is zero, and the blue curve is over a water-lubricant interface, where a non-zero velocity is measured. From these velocity profiles, the local slip length $b$ is calculated as the ratio between the slip velocity and the velocity gradient,

$$b = \frac{U_s}{(\partial u/\partial z)|_{z=0}}. \quad (16)$$

Figure 12(c) shows the slip velocities (left panel) and the corresponding slip lengths (right panel). The liquid interface forms a curved meniscus that is deflected into the groove. The curvature of the meniscus, extracted from the B-scan, provides insight into the relationship between interface curvature and slippage.

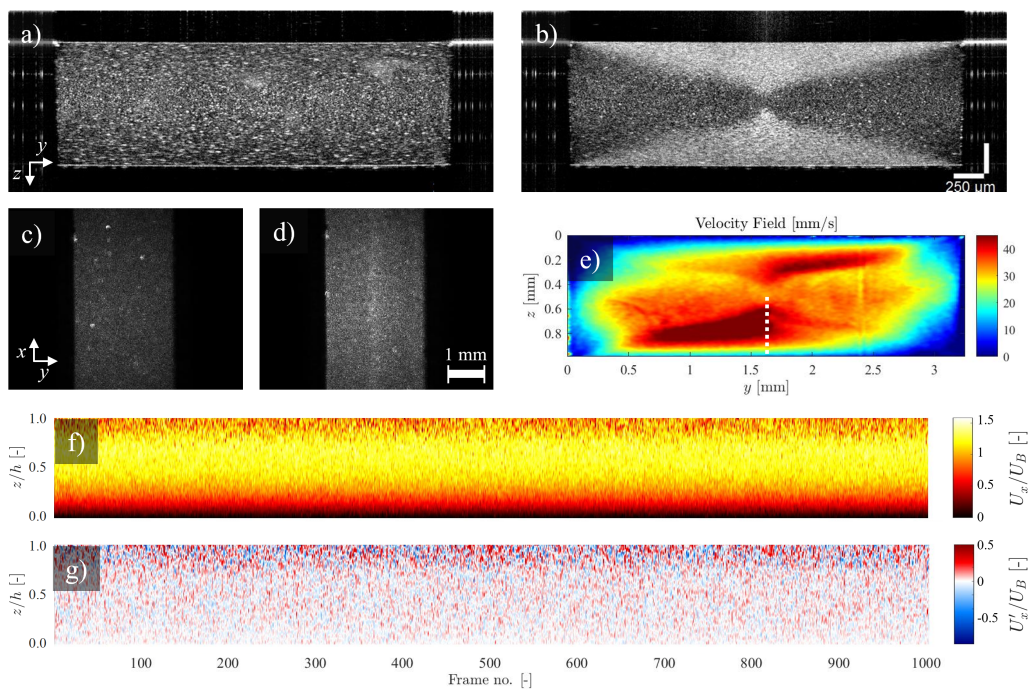

Fig. 13 Flow of polyacrylamide (PAA) in a rectangular duct. a) Intensity measurement on the fluid at rest. b) Intensity measurement on the flow at $0.4\,\mathrm{mL\,min^{-1}}$. c) Polarized Light Imaging on the fluid at rest. d) Polarized Light Imaging on the flow at $0.4\,\mathrm{mL\,min^{-1}}$. e) Mean velocity field on the duct cross-section. f) Time-resolved half-height velocity profile spacetime. g) Time-resolved half-height fluctuation velocity profile spacetime.

### 4.4 Shear-Induced Structure Detection

Since the OCT intensity signal arises from the back-scattering properties of the material, structural heterogeneities can be detected if they cause sufficiently large refractive index variations. In this section, we demonstrate an example of the detection of Shear-Induced Structure (SIS) in single-phase polymeric fluids. Here, local shear enhances the nematic order within an otherwise entangled phase at rest[62,63].

Figure 13(a) shows a B-scan (cross-sectional intensity) of a highly concentrated aqueous solution of high molecular weight ($M_W > 15 \times 10^6$ Da) polyacrylamide in quiescence. Since the fluid sample is transparent – and thereby undetectable by OCT – a rheologically insignificant amount of Rhodamine dye (0.05% wt./vol.) has been added to it, which is seen in Figure 13(a) as bright dots. When we introduce a flow in the channel, it is observed that due to rheologically complex attributes of this viscoelastic fluid, the high shear rate regions of the duct in flow undergo a nematic phase change (Fig. 13b). The bright and dark regions in the OCT intensity shows a distinctive footprint of the so-called *oriented* and *entangled* phases in the polymer network. To capture the effect of orientation of the polymer chains near the high shear rate regions, Polarized Light Imaging (PLI) measurements are performed on the fluid at rest (Fig. 13(c)) and in flow (Fig. 13(d)). The brighter signal obtained from the center line of the top view of the duct indicates higher contribution of the oriented (bright) region integrated along scanning (i.e., $z$) direction.

Figure 13(e) shows the mean velocity distribution. We observe a complex flow due to the interplay between the orientation, modified rheology, and the instantaneous velocity field leads. A locally higher velocity of the fluid in the relatively low viscosity phase in the oriented regions is also visible. The modified local rheology of the medium determines the amplitude of fluctuations, as observed in the mean (Fig. 13(f)) and fluctuation (Fig. 13(g)) space-times. The data highlights that higher temporal variations of possibly elastic instabilities occur in the entangled phase, which retains its initially high elasticity. Note that the time-resolved velocimetry data presented in Figure 13 is derived by binning every 50 consecutive A-Scans into a single data point. None of the signal processing steps outlined in Section 4.1 have been applied here, as the focus of this analysis is to identify the locations of high-fluctuation regions within the flow field, rather than the dynamics of large-scale structures.

### 4.5 Fluid-Structure Interaction

In this example, we demonstrate how OCT can be used for studying the interaction between fluid flows and deformable solids. The lower wall of a millifluidic channel is patterned with a surface composed of regularly placed pillars whose hight corresponds to 60% of the channel height. The pillars, made of a flexible resin, are bent when a laminar flow is imposed in the channel at the flow rates of 20, 50 and $90\,\mathrm{mm^3\,s^{-1}}$. The shear force caused by the flow will bend the pillars, and the extent of the deflection $\delta$ depends on both the force applied and the material properties

Journal Name, [year], [**vol.**],1–17 | 13

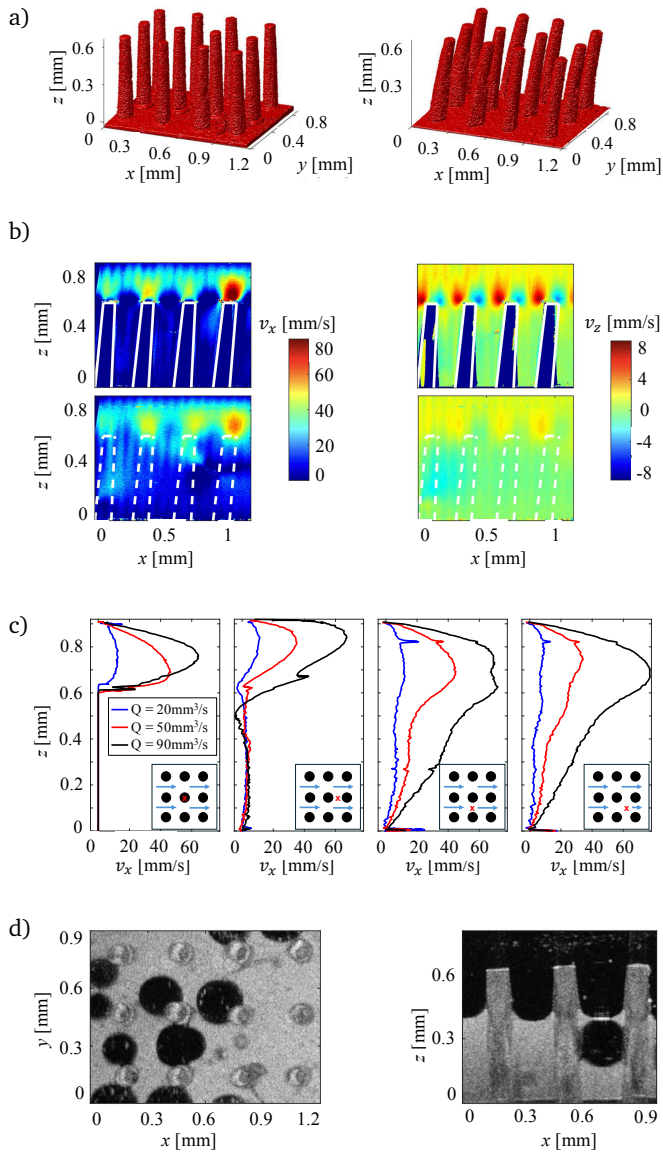

Fig. 14 Study of interaction between flexible pillars and laminar flow. (a) The pillars reconstructed from a 3D scan with OCT are straight in absents of flow (left) and bent at the flow rate of $Q = 70\,\mathrm{mm}^3\,\mathrm{s}^{-1}$ (right). (b) Velocity field in the channel for $Q = 50\,\mathrm{mm}^3\,\mathrm{s}^{-1}$ where the streamwise component is shown in the tiles on the left and the wall-normal component in the tiles on the right. Where the pillars are represented with solid white lines the measurement plane crosses a line of pillars whereas if the pillars are shown as dashed lines, the plane is in between them. (c) Streamwise component of the velocity at different locations and for different flow rates. The locations are represented with a red cross on the map at the right bottom of each plot. (d) 2D cut of the system with pure silicone oil in black and pure milk in white, viewed from above (left) and viewed from the side (right). Bubbles of oil are present between pillars. Figure adapted from Ricard[64].

of the pillars. At equilibrium, the bending of the pillars is determined by balancing the fluid shear force and the pillar stiffness. The non-dimensional deflection $\delta$ at equilibrium of pillar with the diameter $d$ can be written as

$$B = \frac{\delta}{d} = \frac{128\eta Q L^3 p^2}{\pi (H-L)^2 D E d^5} \qquad (17)$$

where $\eta$ is the fluid viscocity, $Q$ the flow rate, $L$ is the pillars' height, $E$ the Young modulus of the pillars and $H$ and $D$ are the height and width of the channel, respectively.

Using a 3D C-scan, we measure the deflection of the structure from its resting condition to estimates its bending stiffness. Figure 14(a) shows how the pillars bend in response to fluid flow, and quantitative measurements can be used to estimate $E$ and the bending modulus $B$. Velocity measurements are carried out using D-OCT at two different inclinations to measure the streamwise and wall-normal components of the velocity field around the pillars (Fig. 14(b) and (c)). These measurements confirm the presence of Poiseuille-type flow above the pillars and provide insights into how the fluid followed the curve of the pillars as they bend. As final demonstration, we use both miscible (glycerol and milk) and immiscible (silicone oil and milk) fluids in static and dynamic conditions. In the glycerol-milk system, milk gradually displace glycerol from above, and with OCT scans the interface between the two fluids are visualized as they mixed. In contrast, the silicone oil-milk system, due to the immiscibility and different densities, exhibited a more complex interaction, with milk displacing oil from below, and oil bubbles being trapped between the pillars, as shown in Figure 14(d).

### 4.6 Fluid Fluid Interaction

In this final example, we use OCT to capture the movement of oscillating drops in a laminar flow.

We consider a millifluidic channel where the bottom wall is engraved with longitudinal grooves exposed to a steady laminar flow ($3.5\,\mathrm{mL\,min^{-1}}$) of a milk-water mixture. The cross section of the grooves have dimensions $243 \times 376\,\mu\mathrm{m}^2$. Lubricant oil (hexadecane) infuses the grooves and eventually elongated drops are formed due to capillary instabilities at the water-oil interface. The front of the drops is pinned to the bottom of the grooves because of the high contact angle hysteresis. The water-lubricant interface of the confined droplets are captured under dynamic conditions and the height maps are derived from the intensity data. The compact design of the OCT device allows it to be directly integrated with the small-scale microfluidic setup, enabling volume acquisitions during flow experiments. These observations reveal the shape alterations of lubricant droplets under laminar flow. When the duct is filled but no flow is present, the drops' interface remain flat along the streamwise direction (Fig. 15(a)). However, under stationary flow conditions, the interfaces deform, protruding over the adjacent ridges (Fig. 15(b)). Panels (c) and (d) show the 2D projections of the surfaces highlighting the vertical deformation in several spanwise positions inside the central groove of the analyzed volume.

In the same experiment, a series of intensity acquisitions of the vertical plane along a single drop was captured at a frequency of $5\,\mathrm{Hz}$ over approximately 7 minutes. As shown in the image series (see video in the SI), the oscillating motion of the interface is clearly resolved. This demonstrates the capacity of using OCT for capturing multi-phase dynamics. However, there are some limitations to consider. The optical field of view is constrained to the millimeter scale, depending on the lens used, and the temporal resolution $t$ depends on the lateral dimension of the field of view, number of pixels N, and the scanning frequency $f_S$ as $t = \frac{N}{f_S}$. For the acquisition of the B-scan, one should add $2\,\text{ms}$ which is approximately the time needed for the probe to move back at the initial position after each A-scan. This temporal limitation should be compared to the characteristic time scale of the phenomenon under study to find the best settings.

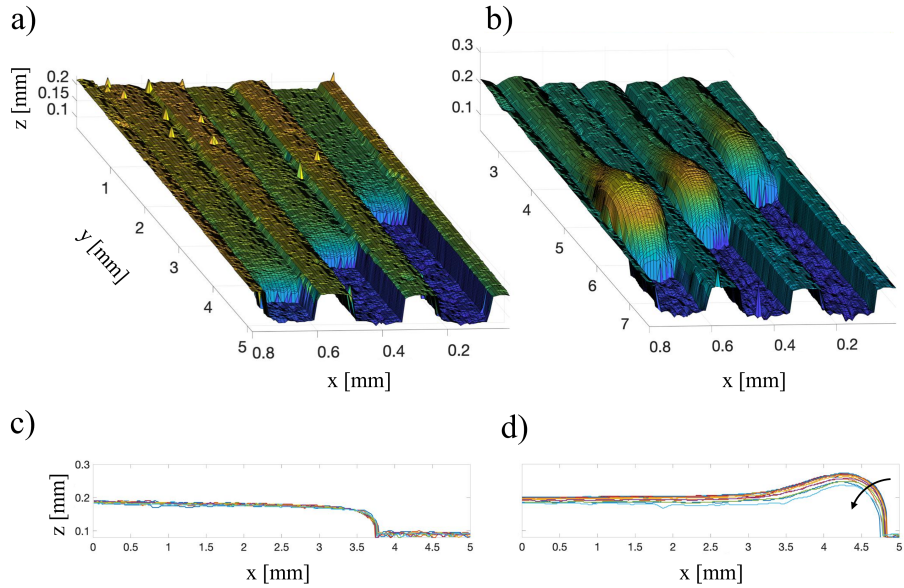

Fig. 15 Topological 3D scans of hexadecane drops confined in parallel grooves. a) The interface liquid-solid and liquid-liquid is captured in absence of flow and b) with a flow rate of $3\,\text{mL}\,\text{min}^{-1}$. c) The interface profiles of the central drop at different spanwise position are plotted for the steady and d) in the dynamic case. The arrow in (d) shows the direction of the spanwise position of slices towards the groove side-walls.

## 5 Notes to Practitioners

In this section, we share a set of practical guidelines drawn from our experience with OCT measurements. While these guidelines reflect general principles for operating OCT systems, case-specific adjustments may be necessary to achieve optimal results.

1. Avoid autocorrelation noise – visible as horizontal bright lines – by increasing the A-scan rate or avoiding reflective interfaces that are perpendicular to the light beam.

2. A useful range of inclination angles for D-OCT measurements is between $3°$ and $9°$, allowing for a high range of velocities over an almost vertical cross-section.

3. The bottom surface of a flow-cell should be matte and opaque to minimize optical artefacts.

4. The thickness of the optical access window – usually with a refractive index larger than water – should be limited to a few millimeters. Thick windows artificially increase the path length, potentially requiring modification of the reference arm of the OCT.

5. Spectral averaging – repeating and averaging each A-scan in D-OCT measurements – decreases the effective A-scan rate. For a given target scan rate, a higher acquisition rate with the corresponding averaging appears to yield better results.

6. Excessive noise in Doppler measurements may be caused by insufficient overlap between adjacent A-scans. This overlap should be larger than $50\%$.

7. Velocity distributions that are close to the limits of the velocity range may wrap to a negative Doppler phase.

## 6 Conclusions and Outlook

Optical Coherence Tomography (OCT) has emerged as a powerful tool for characterizing fluids, structures and their interactions in soft matter problems. Its unique ability to capture mesoscopic features in both space and time has filled an instrumental gap, enabling researchers to investigate confined and dynamic systems. The short acquisition times, especially compared to confocal laser scanning microscopy, make it possible to capture changes in three-dimensional structures on a scale of minutes. Limiting the acquisition to 2-dimensional B-scans, even faster changes can be measured, enabling for example contactless creep tests. Doppler-OCT can be used to obtain accurate velocity profiles, especially close to walls, where other measurement techniques often struggle.

While OCT has been employed in a variety of soft matter contexts, its potential remains largely untapped in several areas such as active matter, microfluidics, and bioengineering. We anticipate that further developments will push the boundaries of what can be measured, such as real-time polarized light imaging and probing for micro-structure anisotropies. However, there are still challenges that need to be overcome to further broaden the OCT applications. Achieving quantitative measurements in transparent media requires the addition of contrasting agents, which is largely unexplored. This is particularly, challenging in interfacial flows, where the adsorption of the seeding particles can modify the interfacial physics, for example through Marangoni-stress. Currently, Doppler-OCT measurements require some *a-priori* knowledge to correctly scale the velocities as in equation 11. Further research in the dependency of this correction factor on environmental parameters would significantly expand the range of applications of Doppler-OCT.

In this review, we have outlined the technical principles of OCT, provided hands-on tutorials to encourage adoption, and showcased its versatility in applications ranging from time-resolved velocimetry to fluid-structure interactions and interface tracking. We hope that the tutorials and showcases in this work increase the accessibility of OCT, contribute to a broader user base and encourage novel applications.

## Author Contributions

The paper was written by KA, CW, SS with feedback OT, FL and SB.

## Conflicts of interest

There are no conflicts to declare.

## Data availability

Since this is a tutorial review, the data presented has been published by the authors according to the references or prepared specifically for the purpose of this tutorial. In the latter case, the data can be reproduced using the Python scripts. The Python scripts for this article can be found at this repository https://github.com/Fluids-Surfaces-Group/

## Acknowledgements

We would like to thank Christophe Brouzet and Guillaume Ricard for figure 14 and the associated work. CW, SB and SS gratefully acknowledges the support of European Research Council (ERC) through project (CoG-101088639 LUBFLOW). KA, FL gratefully acknowledge the support of the Marie Skłodowska-Curie grant agreement No. 955605 YIELDGAP. OT gratefully acknowledges the support of ERC (StG-852529 MUCUS).

## Notes and references

# Supplementary Information :

# Optical Coherence Tomography in Soft Matter

Kasra Amini[a,‡], Cornelius Wittig[a,‡], Sofia Saoncella[a], Outi Tammisola[b], Fredrik Lundell[a], and Shervin Bagheri[a]

## 1 softOCT – A Python package

In this tutorial review, we use the Python package *softOCT* to provide the necessary functions and code examples. *softOCT* provides several functions that can be used to extract information from OCT data. It includes functionality for detecting surfaces and interfaces, isolating phases in multiphase problems, and Doppler-OCT velocimetry. These functions assume that the data is supplied in the form of .tiff image stacks or otherwise converted to numpy ndarrays of intensity / phase.

### 1.1 Example 1 – Simple thresholding

This example is used in Section 3.1.1. An oil droplet is trapped in a milk-filled microchannel. The milk scatters light, whereas the oil appears transparent. This examples shows how to load and binarize OCT intensity measurements, if a clear separation between the foreground and background exists. Finally, the interface between oil and milk is detected and plotted.

### 1.2 Example 2 – Complex thresholding

This example is used in Section 3.1.2. A biofilm, i.e. bacteria suspended in a polymer matrix, grows on a smooth wall. This measurement contains a large field of view of $1 \times 1 \, \text{cm}^2$. This field of view is composed of many A-scans. These A-scans are acquired by traversing the OCT-beam over the area using two galvo-mirrors. Since these two mirrors cannot occupy the same position, optical aberrations are introduced that increase with increasing distance from the center of the field of view. These aberrations manifest as a warping of the image, causing a horizontal surface to warp vertically. Most of this warping can be removed via calibration functions. In this example, we remove the remaining image warping by detecting the substratum, i.e. the surface at the bottom of the channel, and trimming any signal beneath it.

The contrast between the biofilm and the surrounding medium is much weaker than in the previous example. It is in fact so weak, that the histogram of the intensities reveals a monomodal distribution. Therefore, classical image thresholding methods, such as Otsu's method, are no longer able to separate the biofilm from the background signal. Here, we use a thresholding function that considers the shape of the distribution of intensities in a typical OCT scan, where these conditions are given. Finally, we calculate the thickness of the biofilm in one part of the scan as an example of one property that may be derived from such a scan.

### 1.3 Example 3 – Particle Detection

This example is used in Section 3.1.4. Since OCT enables the capture of vertical image slices through a channel, it lends itself well to particle-based flow measurements. Here, we show an example, where particles group in certain regions of the flow. However, the same evaluation procedure may be used to obtain images for velocimetry methods like PIV or PTV. Due to the low particle fraction, we again need to use the OCT-adapted thresholding.

### 1.4 Example 4 – Doppler-OCT in 1D

This example is used in Section 3.2.1. In this example, we show how to convert the Doppler-phase measurements into velocity profiles and compare the result to the analytical solution of a laminar channel flow. We also show how it may be necessary to scale the resulting velocities based on a known quantity, such as the flow rate or a known centerline velocity.

---

[a] *FLOW and Fluid Physics Laboratory, Dept. of Engineering Mechanics, KTH, Stockholm SE-100 44, Sweden*

[b] *FLOW and SeRC (Swedish e-Science Research Centre), Dept. of Engineering Mechanics, KTH, Stockholm SE-100 44, Sweden*

‡ KA and CW had equal contributions.

(Email address for correspondence: kasraa@kth.se / wittig@kth.se / shervinb@kth.se)

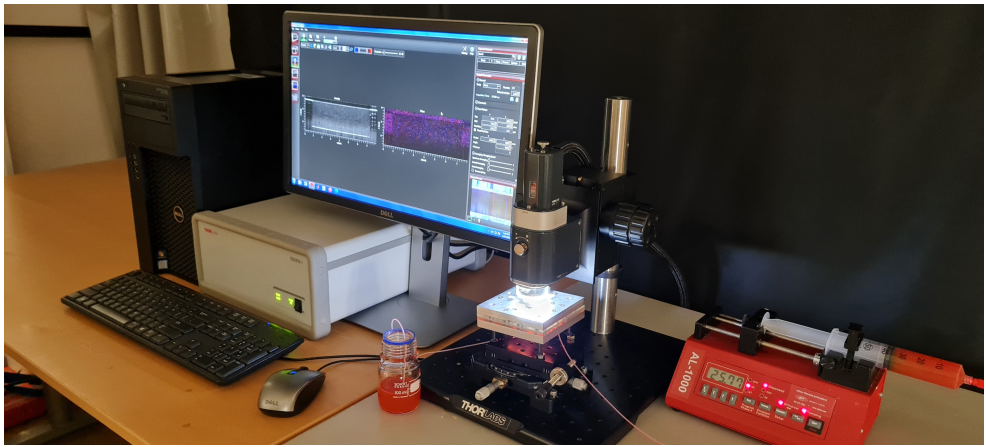

Supplementary Figure 2. The Telesto II Sd-OCT apparatus used for most measurements presented in the current manuscript.

### 1.5 Example 5 – Doppler-OCT in 2D

This example is used in Section 3.2.2. Doppler-OCT measurements may also be acquired along two-dimensional B-scans. Here, we use a set of B-scans to calculate the velocity over the cross-section of a channel. We calculate the evolution of the mean velocity at several locations with increasing sampling time. In this case, our results indicate that at least 250 B-scans are required to obtain a meaningful velocity field. This is caused by the high noise levels in Doppler-OCT measurements, which result in a low precision without degrading the accuracy of the results. Given this data, it is also possible to obtain reliable shear-rates even close to the wall, highlighting one of the advantages of Doppler-OCT measurements. Additionally, this data is used to demonstrate how the wall location may be detected via a maximum in the fluctuations of the intensity measurements (see Section 3.1.3).

## 2 System specifications

As the present manuscript is written in a device-agnostic manner, without any emphasis on the specifics of the OCT apparatus used, with all the accompanying codes solely compatible with a stack of .tiff images, throughout the paper no details and specifications of the used devices were mentioned. However, the data in this tutorial review was acquired using several OCT systems. Here, we present the specifications and limitations of these systems. The technical specifications are listed in Supplementary Table 1.

### 2.1 Thorlabs Telesto II

Most measurements in this tutorial review were conducted with a Spectral Domain Optical Coherence Tomography (Sd-OCT) Telesto II of Thorlabs. The central wavelength of 1310 nm and bandwidth of 270 nm are the nominal specifications of the apparatus. The spatial resolution, therefore, is approximately $2.58\,\mu m$ in depth direction for water. A sample setup as used in e.g. Section 3.1.1 is shown in Supplementary Figure 2. A channel is mounted directly below the OCT probe head.

### 2.2 Thorlabs Ganymede GAN610-SP5

The biofilm measurement in code example 2 was acquired using a Thorlabs Ganymede GAN610-SP5 with a central wavelength of 930 nm and an LSM04 objective lens. In this configuration, an axial resolution of $4.5\,\mu m$ and a lateral resolution of $12\,\mu m$ can be achieved. The wavelength around 900 nm allows for measurements through several millimeters of water. Supplementary Figure 3 shows the corresponding experimental setup. Here, the probe head is mounted to a traverse that is used to move the head to several predefined measurement locations.

## 3 Video

Complementary to Section 4.6, a video showing a drop of lubricant (hexadecane) confined in the grooves of a millifluidic channel is provided at this link. The video has been accelerated by a factor of 30 with respect to real time.

## Notes and references

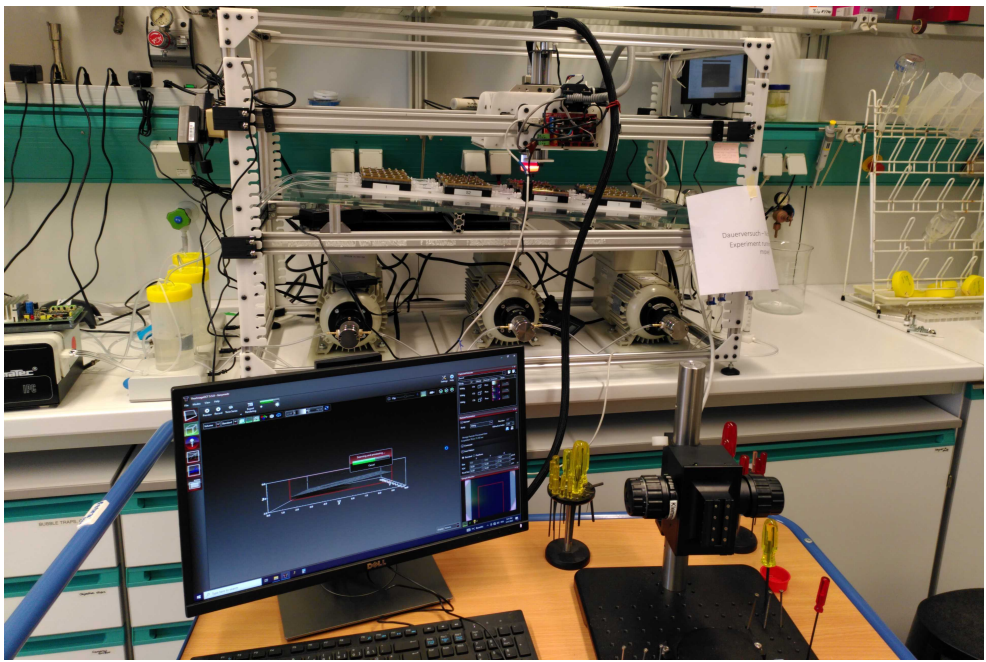

Supplementary Figure 3. The Ganymede Sd-OCT apparatus used to acquire the biofilm sample in section 3.1.2. The probe head is mounted to a traverse.